\documentclass[conference]{IEEEtran}
\IEEEoverridecommandlockouts
\usepackage{cite}
\usepackage{amsmath,amssymb,amsfonts}
\usepackage{algorithmic}
\usepackage{graphicx}
\usepackage{array}
\usepackage{subcaption}
\usepackage{textcomp}
\usepackage{booktabs} 
\def\BibTeX{{\rm B\kern-.05em{\sc i\kern-.025em b}\kern-.08em
    T\kern-.1667em\lower.7ex\hbox{E}\kern-.125emX}}
\begin{document}

\title{Pairwise Judgment Formulation for Semantic Embedding Model in Web Search
}

\author{
\IEEEauthorblockN{Mengze Hong\IEEEauthorrefmark{1}, Di Jiang\IEEEauthorrefmark{1}\IEEEauthorrefmark{2} \thanks{\IEEEauthorrefmark{1}Equal contribution. \quad \IEEEauthorrefmark{2}Corresponding author.}}
\IEEEauthorblockA{Department of Computing\\
Hong Kong Polytechnic University}
\and
\IEEEauthorblockN{Zichang Guo}
\IEEEauthorblockA{Department of Mechanical Engineering\\
Hong Kong Polytechnic University}
\and
\IEEEauthorblockN{Chen Zhang}
\IEEEauthorblockA{Department of Computing\\
Hong Kong Polytechnic University}
}

\maketitle

\begin{abstract}
	Semantic Embedding Models (SEMs) have become a core component in information retrieval and natural language processing due to their ability to model semantic relevance. However, despite its growing applications in search engines, few studies have systematically explored how to construct effective training data for SEMs from large-scale search engine query logs. In this paper, we present a comprehensive analysis of strategies for generating pairwise judgments as SEM training data. An interesting (perhaps surprising) discovery reveals that conventional formulation approaches used in Learning-to-Rank (LTR) are not necessarily optimal for SEM training. Through a large-scale empirical study using query logs and click-through data from a major search engine, we identify effective strategies and demonstrate the advantages of a proposed hybrid heuristic over simpler atomic heuristics. Finally, we provide best practices for SEM training and outline directions for future research.
\end{abstract}


\section{Introduction}
With the growing research interest of extracting latent semantics within text \cite{Jiang2016_latent, Leung2016, hong2025dial, Li2023_mining}, the Semantic Embedding Model (SEM) attracted significant attention from both information retrieval and natural language processing communities \cite{Collobert2011, Wu2014, jiang2015teii}. SEMs are especially useful for web search, drawing on large-scale training data derived from search engine query logs, which contain queries, search results, and various user interaction data \cite{Jiang2016_query, jiang2013panorama, Jiang2016_log, jiang2012g}. In practice, user queries and search results (i.e., titles of retrieved Web pages) are used to construct pairwise training instances. Specifically, for a query $q$, if a retrieved title $d^+$ is preferred over another title $d^-$, this preference is formulated as a pairwise judgment $d^+ > d^-$. SEMs are then trained to increase the similarity between $(q, d^+)$ while decreasing the similarity between $(q, d^-)$.

The problem of deriving pairwise judgments from user click-through behavior has been extensively studied in the field of pairwise Learning-to-Rank (LTR) \cite{Jiang2011, Joachims2007, Radlinski2005, Chapelle2009}. The core idea of these heuristics is to reduce noise and mitigate position bias in the query log by assuming that a document clicked in response to the current query is preferred over a document that was examined but not clicked \cite{Jiang2011}. Despite the intensive research in LTR, relatively little work has addressed the formulation of pairwise judgments for embedding-based models such as SEM. This raises the interesting question of whether the well-established heuristics from LTR still apply to SEM.

In this paper, we investigate this problem using the query log from a major commercial search engine and propose a series of strategies to formulate pairwise judgments. Through extensive experiments, we quantitatively evaluate these strategies and identify the most effective approaches. One key finding is that \textbf{conventional heuristics for pairwise judgment formulation in LTR are not optimal for training SEMs}. This necessitates specialized strategies to generate high-quality training data, a valuable direction for practitioners to improve the performance of embedding-based models. The main contributions of this paper are summarized as follows:

\begin{enumerate}
    \item We provide a detailed methodology for constructing and deploying SEM in real-world Web search scenarios, bridging the gap between theoretical models and practical application;
    \item We present the first in-depth study to propose and rigorously evaluate both atomic and hybrid strategies for formulating pairwise judgments specifically for training embedding-based SEMs;
    \item We identify best practices for training SEM and offer insights into how these strategies differ fundamentally from conventional LTR training approaches.
\end{enumerate}

\section{Related Work}
\label{sec:related work}

The formulation of pairwise judgments has been extensively studied in the context of pairwise Learning-to-Rank (LTR). Radlinski et al. \cite{Radlinski2005} proposed a framework for learning ranked retrieval functions by deriving pairwise preferences from sequences of user queries. Joachims et al. \cite{Joachims2007} investigated the reliability of implicit feedback from click-through data in Web search, showing that user click behavior can provide reasonably accurate preference signals, particularly when comparing documents returned for the same query. These studies form the foundation for deriving pairwise judgments from query logs and have inspired heuristic strategies to mitigate noise and position bias in LTR training data.

Modeling users' browsing patterns is commonly addressed through click models. Chapelle et al. \cite{Chapelle2009} considered click logs as an important source of implicit feedback and proposed a Dynamic Bayesian Network to provide unbiased relevance estimates from these logs. Shen et al. \cite{Shen2012} introduced a personalized click model to capture user-specific click preferences, extending tensor factorization techniques from a collaborative filtering perspective. Chen et al. \cite{Chen2012} proposed a Noise-Aware Click Model that explicitly accounts for varying noise levels in user clicks. Despite their differences, most click models share a key assumption: \textbf{users examine search results sequentially from top to bottom and click on items they perceive as relevant}. This sequential browsing assumption aligns closely with the findings in \cite{Joachims2007}, highlighting the connection between click modeling and pairwise preference formulation.

Building on these insights from LTR and click modeling, recent work has increasingly focused on learning semantic embeddings that capture query-document relevance. Huang et al. \cite{Huang2013} proposed a deep structured semantic model that projects queries and documents into a shared low-dimensional space, where relevance can be measured as the distance between them, and is trained by maximizing the likelihood of clicked documents using click-through data. Shen et al. \cite{Shen2014} extended this approach with a convolutional-pooling structure over word sequences to learn richer semantic representations. The Semantic Embedding Model (SEM) \cite{Wu2014} further improves efficiency by adopting a pairwise training paradigm with hinge loss instead of softmax-based loss, avoiding backpropagation over every training instance. Despite the success of SEM in the search engine industry \cite{Li2021_recalling, Wang2020_search_engine}, there has been surprisingly little work on deriving optimal training data for them. To the best of our knowledge, this study represents the first effort to systematically explore strategies for formulating pairwise judgments for training SEM.

\begin{figure}
    \centering
    \includegraphics[width=1\linewidth]{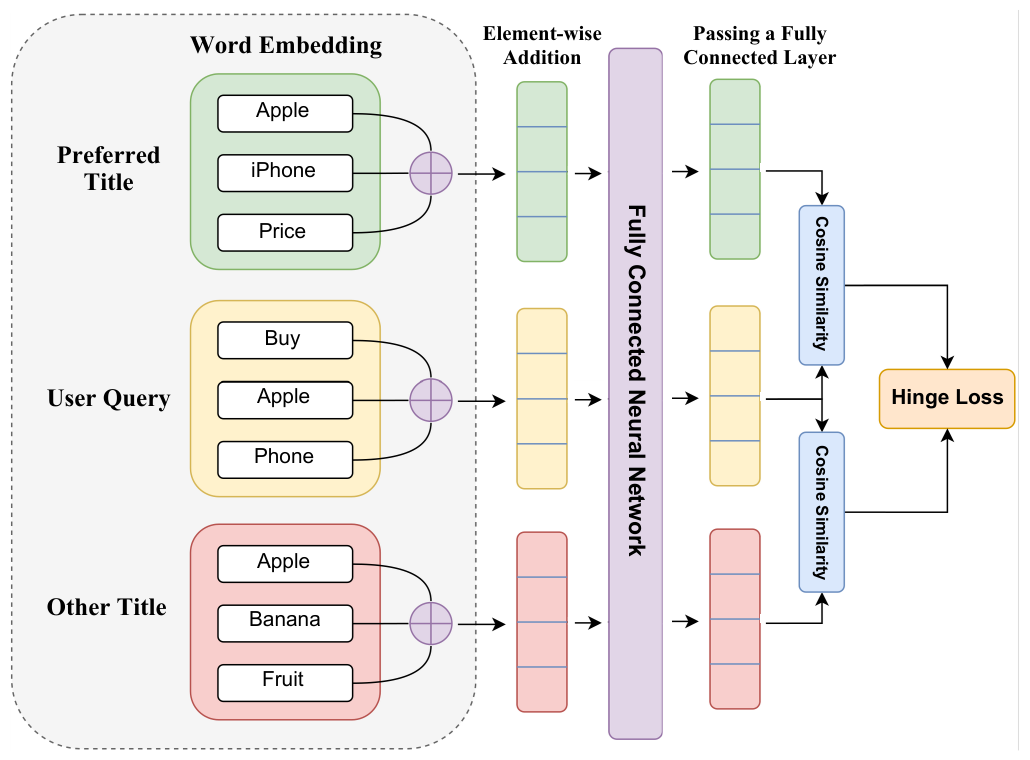}
    \caption{Semantic Embedding Model Architecture}
    \label{fig:sem}
    \end{figure}

\section{Semantic Embedding Model for Web Search}
\label{sec:prerequisites}


\subsection{Architecture of SEM for Web Search} 
We first describe the SEM architecture in the context of Web information retrieval. During training, SEM employs a hinge loss $L$ to optimize the model for distinguishing relevant from non-relevant documents:  

\begin{equation}
L = \frac{1}{m} \sum_{i=1}^{m} \left( \cos \langle f(q_{i}), f(d^+_{i}) \rangle - \cos \langle f(q_{i}), f(d^-_{i}) \rangle \right),
\end{equation}

\noindent where $m$ is the number of training instances, $\cos$ indicates cosine similarity, and $f(\cdot)$ is the function mapping a query or Web page title to its semantic embedding.

As shown in Figure \ref{fig:sem}, the first layer consists of word embeddings. An intermediate representation of the query is obtained by element-wise addition of the word embeddings. Formally, let $x$ denote the input word embedding, $j = 1, \dots, N$ represent the term indices of the query, $h$ be the intermediate query representation, and $i$ the element index of $h$. Then,

\begin{equation}
h_{i}=\sum_{j=0}^{N}x_{i},
\end{equation}

\begin{equation}
g_{i}=softsign(h_{i})=\frac{h_{i}}{1+\vert h_{i}\vert}.    
\end{equation}

The intermediate representation is subsequently processed through fully connected neural network layers to capture higher-order interactions among query terms and produce the final embedding. Let $O$ denote this final embedding, $W$ the weight matrix of the fully connected layer, and $b$ the bias term, such that

\begin{equation}
    O = Wh + b.
\end{equation}

Based on the final embeddings of the query and the title of retrieved document, denoted as $O_q$ and $O_d$, we compute their cosine similarity as follows:

\[
\cos(O_q, O_d) = \frac{O_q^T O_d}{\|O_q\| \|O_d\|}.
\]

\noindent This similarity score can be directly used for ranking or as a feature in more sophisticated ranking algorithms.

\subsection{Optimization} 

The neural network parameters and the word embeddings are updated by conventional backpropagation. The SEM is trained using stochastic gradient descent. Let $\Lambda$ be the parameters and $\Delta = \cos \langle f(q_i), f(d) \rangle$, they are updated as follows:

\begin{equation}
 \Lambda_{t}=\Lambda_{t-1}-\gamma_{t}\frac{\partial\Delta}{\partial \Lambda_{t-1}}, 
\end{equation}

\noindent where $\Lambda_t$ and $\Lambda_{t-1}$ are the model parameters at $t^{th}$ iteration and $(t - 1)^{th}$ iteration respectively, and $\gamma_{t}$ is the learning rate at $t^{th}$ iteration. This process is applied to all training instances and repeated for several iterations until convergence is achieved. The gradient of the model parameters is derived as follows:

\begin{equation}
 \frac{\partial\Delta}{\partial \Lambda_{t-1}}=\frac{\partial\cos(O_{q},O_{d+})}{\partial \Lambda_{t-1}}-\frac{\partial\cos(O_{q},O_{d-})}{\partial \Lambda_{t-1}}.    
\end{equation}

To simplify the notation of calculating the derivatives of $W$, we let $d$ denote $d^+$ and $d^-$, and we let $a$, $b$, $c$ be $O^{T}_{q}O_d$, $\frac{1}{\|Q_q\|}$, and $\frac{1}{\|Q_d\|}$, respectively. Then, we can compute $\frac{\partial\Delta}{\partial W_q}$ and $\frac{\partial\Delta}{\partial W_d}$ by using the following formulas:

\begin{equation}
 {\frac{\partial\mathrm{cos}({O}_{q},{O}_{d})}{\partial W_{q}}}={\frac{\partial}{\partial W_{q}}}{\frac{{O}_{q}^{T}{O}_{d}}{\| {O}_{q} \| \| {O}_{d} \|}}=\delta_{{O}_{q}}^{(q,d)}h_{d}^{T},    
\end{equation}

\begin{equation}
 {\frac{\partial\mathrm{cos}({O}_{q},{O}_{d})}{\partial W_{d}}}={\frac{\partial}{\partial W_{d}}}{\frac{{O}_{q}^{T}{O}_{d}}{\| {O}_{q} \| \| {O}_{d} \|}}=\delta_{{O}_{d}}^{(q,d)}h_{d}^{T},    
\end{equation}

\noindent where $\delta_{{O}_{q}}^{(q,d)} = bcO_d - acb^3O_q$, and $\delta_{{O}_{d}}^{(q,d)} = bcO_q - acb^3O_d$. Similarly, we can compute the gradient of the intermediate representation $\partial\Delta / \partial h$, and then obtain the gradient of the element-wise addition result $\partial\Delta / \partial v$. With the softsign function in our model, each $\delta$ in the element-wise addition result can be calculated as follows:

\begin{figure*}[!t]
    \centering
    \begin{subfigure}[b]{0.3\linewidth}
        \centering
        \includegraphics[width=\linewidth]{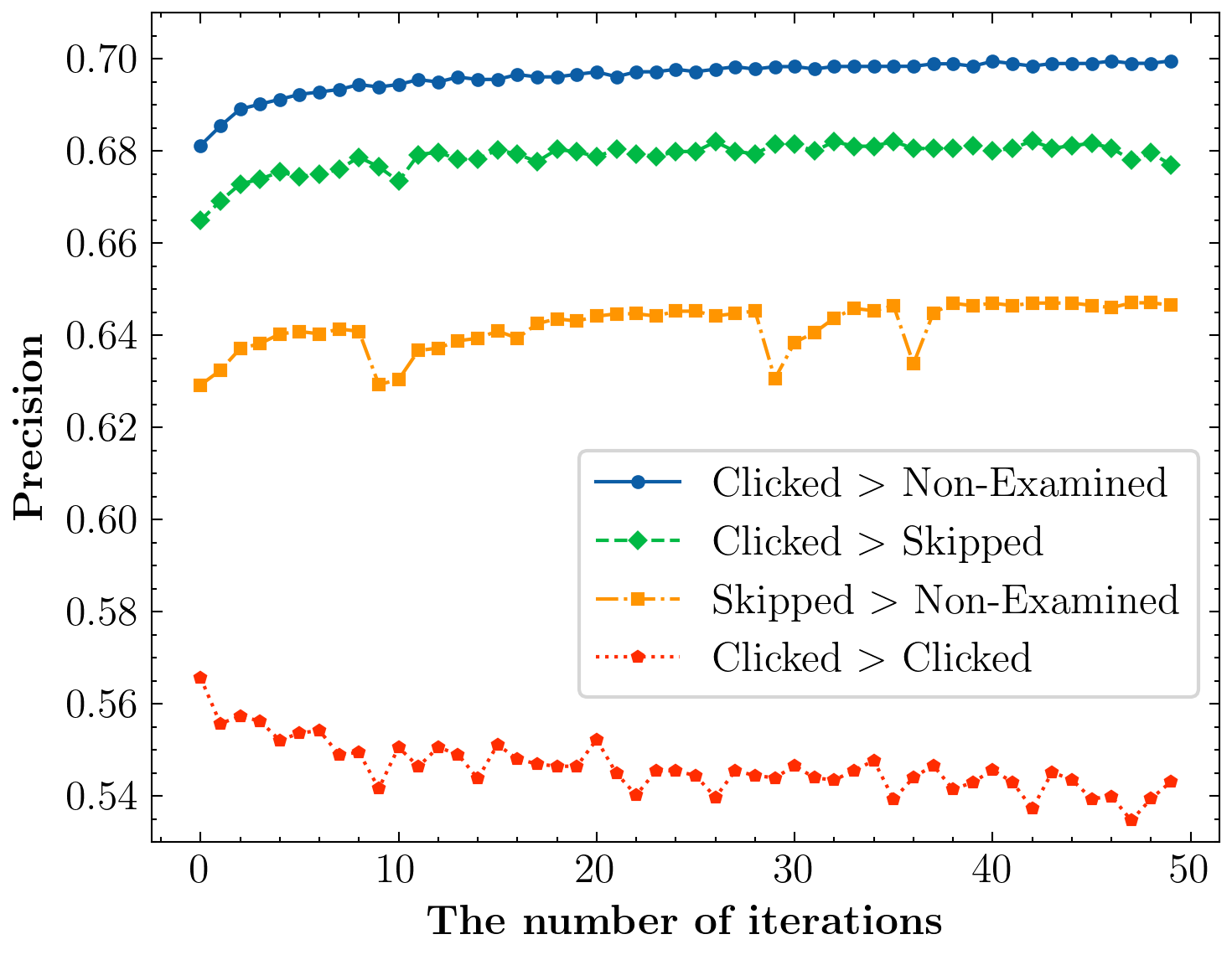}
        \caption{Atomic Strategies: Test-1}
        \label{fig:atomic test-1}
    \end{subfigure}
    \hfill
    \begin{subfigure}[b]{0.3\linewidth}
        \centering
        \includegraphics[width=\linewidth]{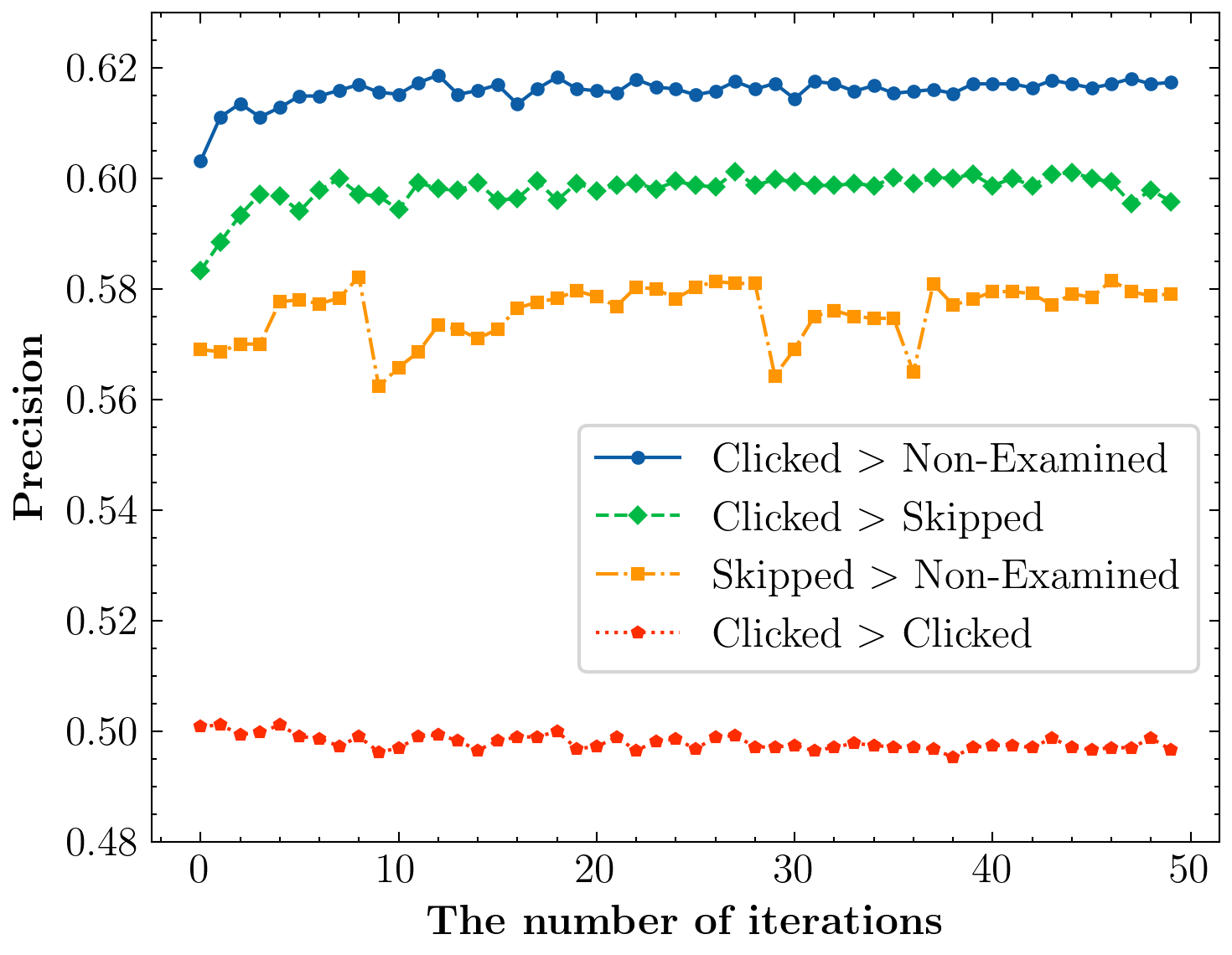}
        \caption{Atomic Strategies: Test-2}
        \label{fig:atomic test-2}
    \end{subfigure}
    \hfill
    \begin{subfigure}[b]{0.3\linewidth}
        \centering
        \includegraphics[width=\linewidth]{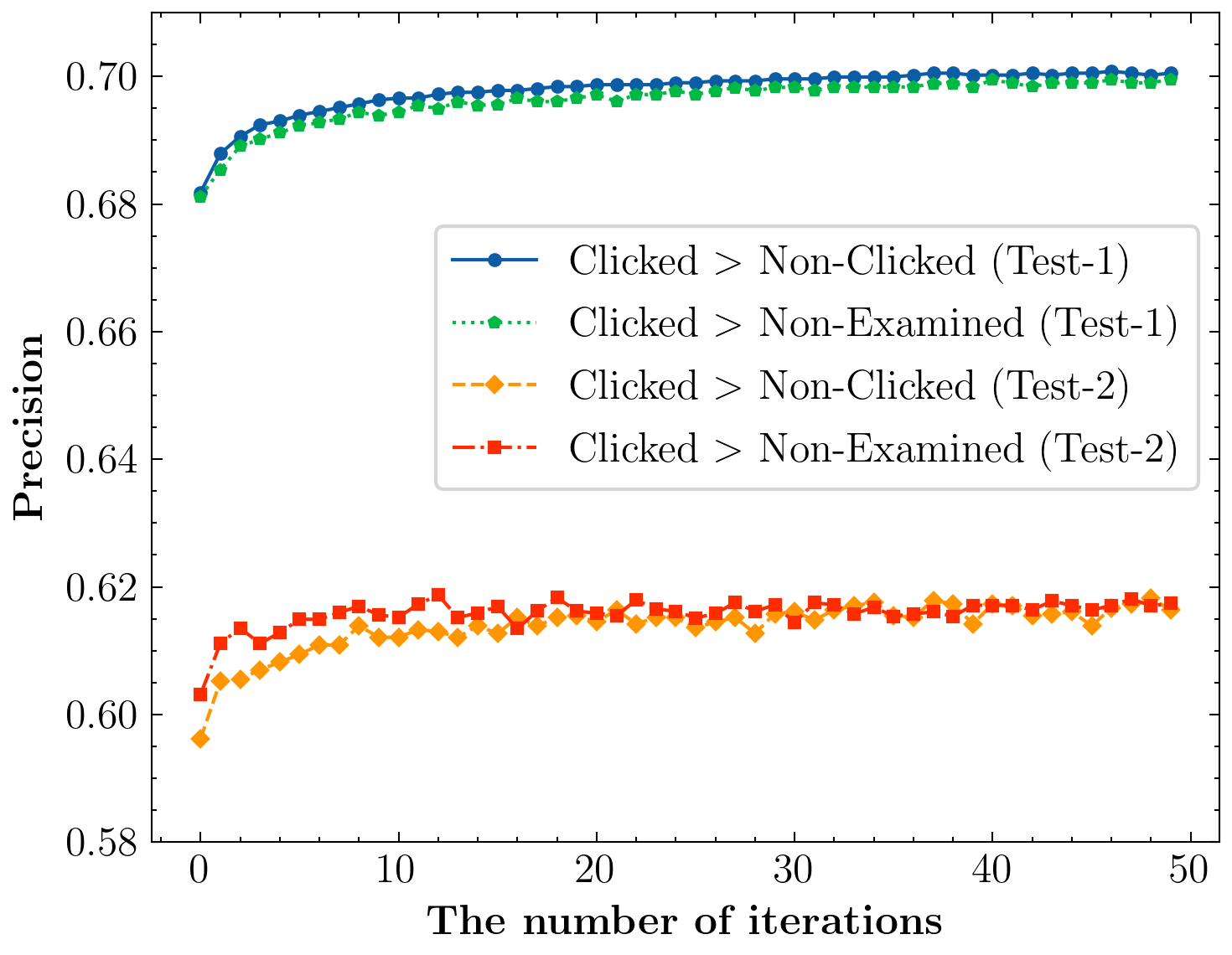}
        \caption{Hybrid Strategy: Test-1 and Test-2}
        \label{fig:hybrid}
    \end{subfigure}
    \caption{Performance of Pairwise Judgments Formed by Various Strategies for Training Semantic Embedding Models}
    \label{fig:result}

\end{figure*}

\begin{equation}
 \delta_{v_{q}}^{(q,d)}=\frac{1}{(1+|V_{q}|)^{2}}\circ W_{q}^{T}\delta_{{O}_{q}}^{(q,d)},    
\end{equation}

\begin{equation}
 \delta_{v_{d}}^{(q,d)}=\frac{1}{(1+|V_{d}|)^{2}}\circ W_{d}^{T}\delta_{O_{d}}^{(q,d)}.    
\end{equation}

\noindent The operator $\circ$ represents element-wise multiplication in the above two formulas. Finally, we backpropagate the gradients of the element-wise addition results, $\delta_{v_q}^{(q,d)}$ and $\delta_{v_d}^{(q,d)}$, to the word embeddings of the query and the document.

\section{Experimental Setup}
\label{sec:setup}

The Web search results of a query can be categorized into three groups based on the user's click signals \cite{Joachims2005}:
\vspace{0.5em}
\begin{enumerate}
    \item \textit{\textbf{Clicked}}: Results that were clicked by users.
    \vspace{0.5em}
    \item \textit{\textbf{Skipped}}: Results ranked above a clicked result that were examined but not clicked.
    \vspace{0.5em}
    \item \textit{\textbf{Non-Examined}}: Results ranked below all clicked results, likely never seen by users.
\end{enumerate}
\vspace{0.5em}

In our experiments, model performance is evaluated using two metrics. The first metric assesses whether the model can effectively predict future clicks by assigning higher scores to clicked results than to non-clicked ones. To evaluate this, we construct a testing dataset, \textit{\textbf{Test-1}}, which contains 23,000,000 pairwise judgments derived from large-scale search query logs. For each query and its top ten results, each pairwise judgment in Test-1 consists of a randomly selected clicked result and a randomly selected non-clicked result.

The second metric evaluates whether the model's results align with human judgment. For this purpose, we use a manually curated testing dataset, \textit{\textbf{Test-2}}, which contains 530,000 pairwise judgments annotated by human experts.

\section{Atomic Strategies}
\label{sec:atomic}

In learning-to-rank (LTR), it is widely accepted that the relative preferences of clicked documents over skipped ones are reasonably reliable \cite{Agichtein2006, Joachims2005}. Motivated by this observation, we propose several strategies to derive pairwise judgments based on a user's ranking preferences, as reflected in their click behavior. These strategies are mutually exclusive and can serve as fundamental building blocks for constructing more complex pairwise judgments.

\begin{enumerate}
    \item \textit{\textbf{Clicked}}  $>$  \textit{\textbf{Skipped}}: This strategy assumes that clicked results are preferred over skipped results, which is the most widely used approach in LTR.
    \item \textit{\textbf{Clicked}}  $>$  \textit{\textbf{Clicked}}: This strategy differentiates clicked results by their click-through rate (CTR) and assumes that a result with a higher CTR is preferred.
    \item \textit{\textbf{Clicked}}  $>$  \textit{\textbf{Non-Examined}}: This strategy assumes that the clicked results are preferred over unseen results.
    \item \textit{\textbf{Skipped}}  $>$  \textit{\textbf{Non-Examined}}: This strategy is rarely applied in LTR since it does not rely on click information; included here for completeness.
\end{enumerate}

Empirically, SEM typically requires several iterations to converge, and we find that 50 iterations are sufficient to obtain a stable model. The experimental results on Test-1 are shown in Figure \ref{fig:result}(a). We observe that the \textit{Clicked} $>$ \textit{Non-Examined} strategy achieves the highest precision, suggesting the most reliable training data for SEM. In contrast, pairwise judgments formed by \textit{Clicked} $>$ \textit{Skipped} and \textit{Skipped} $>$ \textit{Non-Examined} exhibit lower quality, while those formed by \textit{Clicked} $>$ \textit{Clicked} tend to have the lowest quality, due to the lack of meaningful preference information between equally clicked results. 

Interestingly, this result differs from conventional LTR, where the best pairwise judgments are typically derived between skipped and clicked results, i.e., the \textit{Clicked} $>$ \textit{Skipped} strategy \cite{Joachims2007}. Also, the \textit{Clicked} $>$ \textit{Non-Examined} strategy, which is rarely used in LTR, performs best for training SEM, with a significant performance gap compared to the other strategies. We further evaluated the strategies on Test-2, and the results shown in Figure \ref{fig:result}(b) reveal that, although the overall precision on Test-2 is lower, likely due to the increased difficulty of the testing data, the relative performance trends and rankings of the strategies remain consistent with those observed on Test-1.  

In both testing datasets, \textit{Clicked} $>$ \textit{Clicked} is the only strategy that leads to a decrease in precision as more iterations are conducted, indicating that the pairwise judgments derived from this strategy are nearly redundant. Another notable observation is the variation in precision across strategies: \textit{Clicked} $>$ \textit{Non-Examined} exhibits the most stable performance, \textit{Clicked} $>$ \textit{Skipped} shows greater variability, and \textit{Skipped} $>$ \textit{Non-Examined} demonstrates the largest fluctuations. The magnitude of these variations reflects the extent to which training can be occasionally distorted by low-quality training instances.

\section{Hybrid Strategy}
\label{sec:hybrid}
In the previous section, the atomic strategies are investigated, and the \textit{Clicked} $>$ \textit{Non-Examined} is identified to provide the best performance in both tasks. A large performance gap between any two atomic strategies is observed. Intuitively, combining these atomic strategies will not bring better results than \textit{Clicked} $>$ \textit{Non-Examined} since the low-quality training instances will contaminate the result. However, through extensive empirical evaluation, we find that the intuition holds but with one exception, which results in the following hybrid strategy: \textbf{\textit{Clicked} $>$ \textit{Non-Clicked}}, a combination of \textit{Clicked} $>$ \textit{Skipped} and \textit{Clicked} $>$ \textit{Non-Examined}.

\begin{table}[!htbp]
\centering
\caption{Distribution of Pairwise Judgment Dataset}
\label{tab:atomic}
  \begin{tabular}{cc}
    \toprule
    \textbf{Strategy} & \textbf{Percentage} \\
    \midrule
    Clicked  $>$  Clicked & 5.96\%\\
    Clicked  $>$  Skipped & 22.25\%\\
    Skipped  $>$  Non-Examined & 32.92\%\\
    Clicked  $>$  Non-Examined & 38.87\%\\
  \bottomrule
\end{tabular}
\end{table}

Based on the experimental results shown in Figure \ref{fig:result}(c), we observe that \textit{Clicked} $>$ \textit{Non-Clicked} slightly outperforms the best atomic strategy, \textit{Clicked} $>$ \textit{Non-Examined}, on Test-1. To understand the underlying reasons, we present the distribution of the four strategies in Table \ref{tab:atomic}. The statistics indicate that no single strategy covers a majority of the pairs, with \textit{Clicked} $>$ \textit{Skipped} and \textit{Clicked} $>$ \textit{Non-Examined} accounting for only 22.25\% and 38.87\% of the potential training data, respectively. Consequently, relying solely on a single strategy like \textit{Clicked} $>$ \textit{Non-Examined} excludes many reasonably good training instances, limiting opportunities to update the embeddings. Over the long run, the model using the \textit{Clicked} $>$ \textit{Non-Clicked} strategy is exposed to more training instances, resulting in slightly better performance than the atomic strategies. Another possible explanation is that the Test-1 data are primarily derived from the \textit{Clicked} $>$ \textit{Non-Clicked} strategy, which may provide the model with a better fit on this specific test set.


To assess whether the performance of \textit{Clicked} $>$ \textit{Non-Clicked} is exaggerated by the Test-1 data, we evaluate the hybrid strategy on Test-2, where the results differ from those of Test-1. Initially, \textit{Clicked} $>$ \textit{Non-Examined} outperforms \textit{Clicked} $>$ \textit{Non-Clicked}, suggesting that high-quality pairwise judgments provide a better starting point for training the SEM. Over time, however, the performance gap between the two strategies narrows, and they achieve similar performance upon convergence. Nevertheless, the hybrid strategy exposes the model to more training instances and remains a promising approach for consideration in SEM training.


\section{Discussions}

Based on observations from the experimental results, the insights and best practices for pairwise judgment formulation are summarized as follows:

\begin{enumerate}
    \item The conventional LTR strategy for pairwise judgment formulation is not well-suited for training SEM. LTR focuses on learning the weights of features in a ranking function, whereas SEM aims to learn effective representations of the features themselves.
    \item When \textit{Clicked} $>$ \textit{Non-Examined} is applied, incorporating pairwise judgments derived from \textit{Clicked} $>$ \textit{Skipped} to form a hybrid heuristic can slightly improve performance by increasing the diversity of training instances.
    \item The strategy \textit{Clicked} $>$ \textit{Non-Examined}, which is rarely used in LTR, produces the most effective training data for SEM. Using this strategy alone results in a smaller dataset (38.87\%) but achieves performance nearly as good as the hybrid strategy \textit{Clicked} $>$ \textit{Non-Clicked},  which combines both atomic strategies to create a much larger dataset (61.10\%).
\end{enumerate}
These findings highlight the importance of carefully selecting formulation strategies to achieve effective SEM training.

\section{Conclusions}
\label{sec:conclusion}

In this paper, we examine pairwise judgment formulation for the Semantic Embedding Model using query log data from a major search engine. Large-scale experiments compare various strategies, highlighting key differences from traditional pairwise Learning-to-Rank methods. The strategy \textit{Clicked} $>$ \textit{Non-Examined}, although rarely used in LTR, provides the highest-quality training data for SEM, while the hybrid strategy \textit{Clicked} $>$ \textit{Non-Clicked} offers marginal improvements by leveraging greater data diversity. Future work is encouraged to incorporate additional signals into pairwise judgment formulation, explore strategies for SEM variants, and leverage more powerful word embedding models \cite{devlin2019bert} as well as emerging large language models with advanced semantic understanding capabilities \cite{hong-etal-2025-qualbench} to further enhance SEM performance in Web search. Moreover, pairwise judgments hold significant potential for training advanced semantic relevance models, as they provide reliable supervision that can guide superior language models to better capture query-document relationships. This, in turn, can support the development of next-generation search engines and information retrieval systems, enhancing overall user experience \cite{hong-etal-2025-augmenting, lu2025contextualizedtokendiscriminationspeech}.


\bibliographystyle{IEEE_bibliography/IEEEtran}
\bibliography{IEEE_bibliography/reference}

\end{document}